\documentclass[reprint,amsmath,amssymb,braket,aps,prb,longbibliography,footinbib]{revtex4-2} % PRX
\usepackage{graphicx,color,braket,amsmath,dcolumn,bm,float}% Include figure files
\usepackage[colorlinks=true,linkcolor=blue]{hyperref}
\usepackage{dblfnote}

\begin{document}
\title{Exact theory of plasmon reflection and transmission in partially gated two-dimensional system}

\author{I.M. Moiseenko$^{1}$, D.A. Svintsov$^{1}$}
\affiliation{$^{1}$Center for Photonics and 2d Materials, Moscow Institute of Physics and Technology, Dolgoprudny 141700, Russia}
\email{MoiseenkoIM@yandex.ru}

\begin{abstract}

We develop an exact theory of plasmon scattering at the boundary between gated and ungated regions of a two-dimensional electron system (2DES). Using the Wiener–Hopf technique, we derive analytical expressions for the complex reflection and transmission coefficients of plasmons incident from both sides of the interface. The theory fully accounts for evanescent fields at the gate edge and radiative losses into free-space electromagnetic waves. In the non-retarded limit and for small gate–2DES separation, the reflected plasmon dominates the total electric field, while radiative losses are negligible when plasmon scattering. The amplitudes and phases of the reflection and transmission coefficients for plasmons incident from both sides have a complex dependence from 2DES-gate separation and conductivity of 2DES. %We also provide simplified analytic expressions for the factorized dielectric functions that accurately reproduce the reflection and transmission amplitudes over a wide range of parameters. 
Our results provide a rigorous foundation for modeling tunable plasmonic crystals based on 2DES for terahertz detection and modulation.
\end{abstract}
\maketitle
\section{Introduction}
Two-dimensional (2D) plasmons are collective charge density excitations in two-dimensional electron systems (2DES) exhibit strong sub-wavelength confinement combined with efficient electrical tunability via control gates \cite{Stern:1967, Chaplik:1972}. These properties make 2D plasmons potentially applicable in THz optoelectronics, including tunable detectors, sources, phase shifters, and modulators \cite{Knap:2009, moiseenko2025plasmon, Muravev:2023, Allen:2002, Boubanga-Tombet2020}. The main problem hindering application of plasmons is the wavevector mismatch between incident photons and plasmons, which prevents their direct coupling. This issue is typically addressed by placing a periodic grating gate above the 2DES \cite{Allen:2002,aizin2007terahertz}. Such a grating not only serves as an efficient coupling element but also creates a periodic modulation of the electron density and screening in the 2DES, forming a so-called planar plasmonic crystal \cite{Aizin:2012, Kachorovskii:2012, Kachorovskii:2024, Knap:2023}. By analogy with photonic crystals, its spectrum acquires a band structure with allowed and forbidden bands \cite{Bylinkin_tight_binding, Knap:2023} and enables the excitation of different types of plasmonic modes \cite{Kachorovskii:2012, Popov:2015, fateev2021terahertz}. The theoretical description of plasmonic crystals often relies on the Kronig–Penney model \cite{miranda2024topology, Gorbenko_LateralPC}, which accounts for the periodicity of the structure and allows analysis of both the eigenmodes of the system and its response to external electromagnetic excitation \cite{Muravev:2022}.

The structural element of such plasmonic crystals is the boundary between the screened and unscreened regions of the 2DES. Understanding the scattering of a 2D plasmon at such a discontinuity is necessary for describing the electrodynamics of 2D systems \cite{Moiseenko_partly_gated, Sydoruk_gate_edge,Aizin_finite,lee2025ultranarrow}. This problem reduces to determining the complex reflection and transmission coefficients, which define the amplitude and phase acquired by plasmons upon scattering . The latter has proven particularly important, since, as shown in recent works \cite{moiseenko2026unconventional, Accompanying, lee2025ultranarrow}, the reflection of plasmons from the edge of a metal gate is accompanied by a nontrivial phase shift, leading to the emergence of new types of localized states, such as slot plasmons, whose properties are determined precisely by scattering processes at the boundaries.

There are various approaches for describing plasmon scattering at the boundary between a gated and an ungated 2DES, but they are either insufficiently accurate \cite{Rejaei2015} or do not allow, within a single framework, the investigation of all relevant characteristics of the scattered waves — for instance, the plane-wave matching method predicts a trivial reflection phase of 0 or $\pi$ \cite{Aizin:2012}, while in fact it can have a complex dependence on the parameters of the structure, which is manifested in particular in numerical experiments ~\cite{Sydoruk_gate_edge,Siaber2019}.
In this work, we present a comprehensive theoretical study of 2D plasmon scattering at the boundary between gated and ungated regions of 2DES. Using an exact solution of the scattering problem via the Wiener–Hopf method, we derive rigorous analytical expressions for the amplitude and phase of the reflection and transmission coefficients, and show that the key control parameters are the gate-channel distance and the 2DES conductivity.

%\begin{figure}% [ht]
%\center{\includegraphics[width=0.9\linewidth]{}}
%\caption{(a) Schematic illustration of plasmon reflection and transmission at a  gate edge above 2DES.} 
%\label{structure}
%\end{figure}

The derivation of the complex plasmon reflectance at the gate edge, $r$ and transmittance $t$, starts from Maxwell’s equations for the partially gated 2DES. A perfectly conducting gate is located a distance $d$ above a 2DES and occupies the half-plane $x>0$. The electrodynamics of 2DES is characterized by uniform complex surface conductivity per unit square of area $\sigma = \sigma'+ i \sigma''$, where single and double primes distinguish between real and imaginary parts of a complex quantity. In order to focus on the electrodynamic aspect of the problem, we do not present a specific model of the 2DES conductivity $\sigma(\omega)$, using instead the dimensionless conductivity normalized by the  free space impedance $\eta = \sigma Z_0/2$, where $Z_0$ is the free-space impedance equal to $4\pi/c$ in Gaussian units and 377 Ohm in SI units, $c$ is the speed of light. Note that the presented mathematical framework allows the use of various conductivity models such as kinetic ~\cite{Falkovsky2007a}, hydrodynamic \cite{svintsov2018hydrodynamic}, including those taking into account its nonlocality.

%For monochromatic incident wave with frequency $\omega$, the solution of scattering problem depends only on the conductivity at that particular frequency, $\sigma(\omega)$. The dispersion function $\sigma(\omega)$ can be arbitrary, and can be independently determined from kinetic models~\cite{Falkovsky2007a} or spectroscopic experiments~\cite{dahl2007magneto}. Focusing on electrodynamic aspects of the problem, we refrain from any microscopic model of $\sigma$, and use it as a free parameter.} 

\section{Theoretical method}
We begin the derivation of reflection and transmission coefficients for ungated plasmons with the wave equation for vector potentials $\mathbf{A}$. After the Fourier transform with respect to the $x$-coordinate (${{\mathbf{A}}_{\text{q}}}(z)=\int_{-\infty }^{\infty }{\mathbf{A}(x,z){{e}^{-iqx}}dx}$), the wave equation acquires the following form \cite{Moiseenko_partly_gated}:
\begin{multline}
\label{eq-wave_eq}
\kappa(q)^2{{\mathbf{A}}}(z)-\frac{{{\partial }^{2}}{{\mathbf{A}}}(z)}{\partial {{z}^{2}}}=2Z_0[{{\mathbf{J}}_{2\text{D}}}\delta (z)+{{\mathbf{J}}_{\text{g}}}\delta (z-d)]+
\\
+\mathbf{A}_{\text{ext}}(z),
\end{multline}
where ${{\mathbf{J}}_{\text{g}}}$ and ${{\mathbf{J}}_{\text{2d}}}$ are the Fourier harmonics of the current density, $\mathbf{A}_{\text{ext}}$ is the Fourier transform of the external source potential, where $\kappa(q) = \sqrt{q^2-k_0^2}$ is the decay constant of the electromagnetic field in the $z$-direction, $k_{0}=\omega/c$. 

Using the relationship between potentials and the electric field $\mathbf{E}(z)=-i {\bf q} \varphi_{\bf}(z) + i k_0 {\bf A}_{\bf}(z)$, where $\varphi_{\bf}(z)$ is the scalar potential, taking into account the Lorentz gauge and Ohm's law in 2DES ${{\mathbf{J}}_{2\text{d}}}=\sigma {{\mathbf{E}}}(z=0)$, we obtain an expression for the total electric field in the entire space [Suppl.]:

\begin{gather}
\label{Eq_total}
{{E}_{}}\left( q \right)={E}_{\rm inc}{{e}^{-\kappa(q)|z|}}+i\frac{{{J}_{\rm g}}\left( q \right)}{M(q)}\frac{Z_0}{2 k_0}{{e}^{-\kappa(q)|z-d|}},\\
M\left( q \right)=\frac{\varepsilon_u \left( q \right)}{{{\varepsilon }_{g}}\left( q \right)\kappa \left( q \right)},
\end{gather}
where $\varepsilon_u(q)$ and $\varepsilon_g(q)$ are the dielectric functions of the ungated and gated parts of the 2DES, respectively:
\begin{gather}
\varepsilon_u(q) =1+i\eta \frac{\kappa(q)}{{{k}_{0}}},\\
{{\varepsilon }_{g}(q)}=1+i\eta \frac{\kappa(q)}{{{k}_{0}}}\left( 1-{{e}^{-2\kappa(q)d}} \right),
\label{eq-diel_functions}
\end{gather}
where $\eta = \sigma Z_0/2$ is the dimensionless 2DES conductivity normalized by the free space impedance.  The incident ungated plasmon field  is bounded to the left-half space, so its Fourier transform is ${E_{\rm inc}}={iE}_{\rm 0}/({q-q_u)}$ , where $q_u$ is the wave vector satisfying the dispersion of the ungated plasmon  $\varepsilon_u(\pm q_u) = 0$.  

%To solve the plasmon scattering problem, we nullify the true external field source, $E_{\rm ext} \equiv 0$, and the total field as a sum of incident plasma wave and scattered field, ${{E}_{L}}={E_{\rm inc}} + {E_{\rm scat}}$~\cite{Kay_reactance_discontinuity,Alymov_Refraction}. The incident plasmon field ${E_{\rm inc}}$ is bounded to the left-half space, and its real-space representation is ${E}_{\rm inc} = E_0 {{e}^{i{{q}_{u}}x}} \theta \left( -x \right)$, $q_u$ is the wave vector satisfying the ungated plasmon dispersion $\varepsilon_u(q_u) = 0$}.

In the gate plane ($z=d$), the scattered field is limited to the region $x<0$, while the current in the gate is limited to the region $x>0$. This allows us to apply the Wiener-Hopf method \cite{Noble1958MethodsBO} to solve equation (\ref{Eq_total}), which involves forming the left-hand side of the equation from functions that are analytic in the upper half-plane of the complex wave vector q, and the right-hand side from functions that are analytic in the lower complex half-plane.
To do this, it is necessary to factorize all functions in the form $f(q) = f_+(q) f_-(q)$. The ''plus'' and ''minus'' functions $f_{\pm}$ are obtained from the original function $f$ with the Cauchy theorem:
\begin{equation}
{f_{\pm }}\left( q \right)=\exp \left\{ \pm \frac{1}{2\pi i}\int\limits_{-\infty }^{+\infty }{\frac{\ln f \left( u \right)du}{u-\left( q\pm i\delta  \right)}} \right\}.
\label{eq-factorisation}
\end{equation}
Performing the splitting and equating the parts analytic in the upper (+) and lower (-) complex half-planes to zero, we get the solution for the scattered field and plasmon-induced current in the gate ($z=d$):
\begin{gather}
\label{eq-scat-solution}
{{E}_{\rm L}}\left( q \right) = \frac{i{{E}_0}}{q-{q_u}}\frac{{M}_{+}\left( q_u \right)}{{{M}_{+}}\left( q \right)}e^{-\kappa(q_u)d}, \\ 
{{J}_{\rm g}}\left( q \right) = -\frac{i{E_0}}{q-{{q}_u}}\frac{2k_0{M}_{+}({q}_u){M}_{-}(q)}{Z_0}e^{-\kappa(q_u)d}.
\end{gather}
Substituting the gate current density (\ref{eq-scat-solution}) into (\ref{Eq_total}), we obtain an expression for the total field in the 2DES ($z=0$) containing complete information on the behavior of the reflected ungated plasmon and plasmon transmitted under the gate, as well as the emission of the bulk wave into free space and evanescent fields when scattered at the edge of the gate:

\begin{equation}
\label{Eq_2DES}
{{E}_{2des}}\left( q\right)=\frac{i{{E}_0}}{q-{q_u}}\left[ 1-\frac{{M}_{+}\left( q_u \right){{e}^{-2\kappa(q)d}}}{{{M}_{+}}\left( q \right)\varepsilon_{g}(q)} \right].
\end{equation}
\section{Results: incidence from ungated section}
The amplitudes of reflected ungated and transmitted gated plasmon can be singled out from the total field (\ref{Eq_total}) by the residue of ${{E}_{\rm 2des}}\left( q \right)$ at the poles $q=-q_u$ and $q=q_g$ timed by $i$, respectively:

\begin{gather}
r_{uu} E_0 =i\underset{q=-q_u}{\text{Res}}E_{2des}{\rm}(q).\\
t_{ug} E_0 =i\underset{q=q_g}{\text{Res}}E_{2des}{\rm}(q),
\end{gather} 
where $q_g$ is the gated plasmon wave-vector satisfying the dispersion of the gated plasmon ($\varepsilon_{g}(\pm q_g)=0$). After several straightforward transformations, we arrive at
\begin{equation}
\label{eq-ruu}
{r_{uu}}=-\frac{\kappa(-q_u)}{2{q_u}}{{ \frac{{{M}_{+}}\left( {{q}_{u}} \right)^{2}e^{ -2 \kappa(q_u)d}}{{{{\left. \partial \varepsilon_u /\partial q \right|}_{q=-{q_u}}}}}}}.
\end{equation}

\begin{equation}
\label{eq-tug}
{t_{ug}}=\frac{-1}{q_g-q_u}\frac{M_{+}(q_u)}{M_{+}(q_g)}\frac{e^{-[\kappa(q_u)+\kappa(q_g)]d}}{{\left. \partial \varepsilon_g /\partial q \right|}_{q={q_g}}}.
\end{equation}

%\corra{Equations (\ref{eq-ruu}) and (\ref{eq-tug}) for the plasmon reflectance and transmittance at the boundary between gated and ungated regions is one of our central results. It fully accounts for evanescent waves excited near the gate boundary, as well as plasmon radiative losses, i.e. the emission of free-space electromagnetic waves upon scattering.}

Similar reflection and transmission coefficients were obtained in various approaches ~\cite{Siaber2019}, and the simplest form is that of the coefficients obtained by matching the electric fields of plane waves at the boundary of the ungated and gated 2DES:

\begin{equation}
\label{eq-ruu-pw}
{r_{pw}}=\frac{q_g-q_u}{q_g+q_u}.
\end{equation}

\begin{equation}
\label{eq-tug-pw}
{t^{(pw)}(q_i)}=\frac{2q_i}{q_g+q_u},
\end{equation}

where $q_i$ is the wave vector of the incident plasmon, ${t_{ug}^{(pw)}=t^{(pw)}(q_i=q_u)}$ is the transmittance coefficient for plasmons propagating from the open region of 2DES under the gate and in the opposite direction (transmittance from gated to ungated ${t_{gu}^{(pw)}=t^{(pw)}(q_i=q_g)}$).

\begin{figure}[ht]
\center{\includegraphics[width=\linewidth]{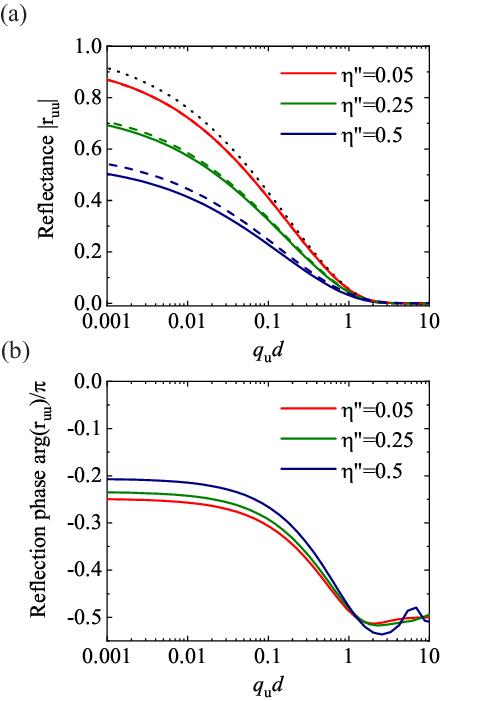}}
\caption{Reflection of the ungated plasmon at a gate edge. Amplitudes of the reflection coefficient $|r_{uu}|$ (solid curves) and $|r_{uu}^{pw}|$ (dotted curve) (a) and its phase, (the phase $arg(r_{uu}^{pw})=0$ is not shown) (b), both plotted as functions of the normalized gate-2DES separation $q_u d$. Different colors correspond to different values of normalized conductivity $\eta$, which is assumed purely imaginary ($\eta' \ll \eta''$). Dashed curves at Fig. (a) correspond show the fully analytical approximation of the reflectance using absolute values of the factorized functions ~\eqref{eq_fact_analyt},~\eqref{eq_fact_analyt_g}.}
\label{Fig-reflectance-ruu}
\end{figure}

\begin{figure}[ht]
\center{\includegraphics[width=1\linewidth]{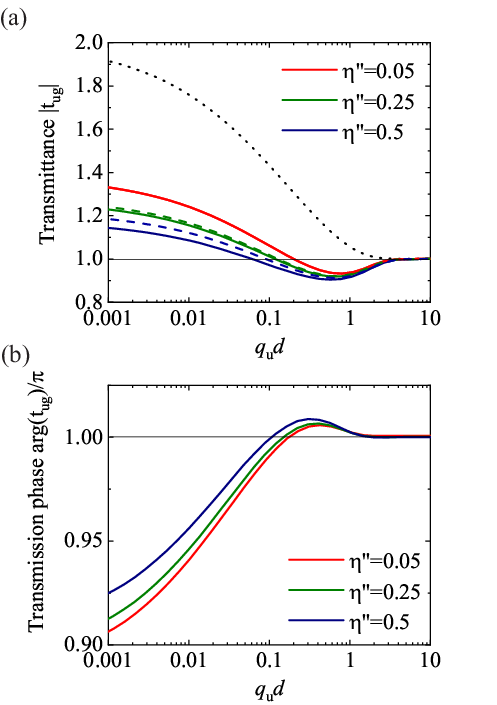}}
\caption{Transmission of the ungated plasmon under the gate. Amplitudes of the transmission coefficient $|t_{ug}|$ (solid curves) and $|t_{ug}^{(pw)}|$ (dotted curve) (a) and its phase (b), both plotted as functions of the normalized gate-2DES separation $q_u d$. Different colors correspond to different values of normalized conductivity $\eta$, which is assumed purely imaginary ($\eta' \ll \eta''$). Dashed curves at Fig. (a) correspond show the fully analytical approximation of the reflectance using absolute values of the factorized functions ~\eqref{eq_fact_analyt},~\eqref{eq_fact_analyt_g}.} 
\label{Fig-transmittance-tug}
\end{figure}

Next, we show that absolute reflectance $|r_{uu}|$ and transmittance $|t_{ug}|$ depend on both the 2DES-gate distance and the dimensionless normalized 2DES conductivity $\eta=\eta'+i\eta''$.  We consider the weakly dissipative limit $\eta' \ll\eta''$ as the most relevant for studying the plasmonic properties of 2DES-based structures. 
Both absolute reflection and transmission are maximized as the gate moves to 2DES, and for a close 2DES-gate separation ($q_ud \ll 1$) they strongly depend on the normalized conductivity $\eta''$ (Fig \ref{Fig-reflectance-ruu}(a) and Fig. \ref{Fig-transmittance-tug}(a)). For a fixed value of $q_ud\ll1$, the coefficients $|r_{uu}|$ and $|t_{ug}|$ grow with decreasing normalized conductivity $\eta''$, which is associated with an increase of plasmon field localization near 2DES ($q_u\gg k_0$), leading to a decrease in radiative losses of plasmons \cite{moiseenko2026unconventional}. Radiative losses is due to the presence of non-zero imaginary part of the dielectric functions (\ref{eq-diel_functions}) for $-k_0<q <k_0$, while outside this range of wavenumbers, both dielectric functions remain real. The ’radiative’ values of $q$ contribute to the modulus of the factorized dielectric functions, while the ’non-radiative’ values do not. 

The absolute transmission coefficient differs from the transmission (\ref{eq-tug-pw}) obtained by the plane wave matching method (dashed curve at Fig. ~\ref{Fig-transmittance-tug}(a)). This is because the expressions (\ref{eq-ruu-pw}) and (\ref{eq-tug-pw}) is obtained taking into account the continuity of currents and potentials at the boundary of the ungated and gated 2DES, although this is valid only at one point along the vertical $z$-axis, namely in the 2DES plane, since the 2DES screening can significantly change the spatial distribution of the potential, while the expression \ref{eq-tug} takes into account the matching of potentials in the entire space. At the same time, absolute reflectance $|r_{uu}^{pw}|$ is in good agreement with $|r_{uu}|$ in non-retarded limit, since the incident and reflected waves are in the same unscreened 2DES. The deterioration of the match between the absolute reflectance coefficient $|r_{uu}|$ with $|r_{uu}^{pw}|$ with increasing $\eta''$ is associated with an grow in radiation losses, which is not taken into account in the plane wave matching method.

% since, unlike the latter, the expressions (\ref{eq-ruu}),(\ref{eq-tug}) take into account the radiative attenuation associated with the emission of bulk waves during plasmon scattering at the gate edge, as well as the presence of evanescent fields near the gate edge.

%Фаза отраженной волны $arg(r_{uu})$ также зависит от степени экранирования 2дэс и в случае ($q_ud \ll 1$) $arg(r)\approx -\pi/4$ в нонретардед лимит (Рис. \ref{Fig-reflectance-ruu}(b)), в то время как метод сшивки плоских волн предсказывает фазу отражения, равную $\pi$. Нетривиальная фаза отражения плазмона от края затвора приводит к неожиданному эффекту изменения правила квантования волнового вектора для стоячих плазмонных мод в щели над двумерной системой, образованной двумя параллельными металлическими затворами, верифицированных экспериментально [PRL Moiseenko, PRL Muraviev]. Фаза коэффициента прохождения $arg(t_{ug})$ также зависит от степени экранирования 2дэс, и может незначительно отличаться от $\pi$ как в большую, так и в меньшую сторону (Рис. \ref{Fig-transmittance-tug}(b)). Интересно, что при $arg(t_{ug})<\pi$ амплитуда прохождения $|t_{ug}|>1$, в то время как $arg(t_{ug})>pi$ имеем $|t_{ug}|<1$. Последняя закономерность сохраняется вплоть до $q_ud \approx 1$ (см. Рис \ref{Fig-transmittance-tug}(a),(b))).

The phase of the reflection coefficient $\arg(r_{uu})$ also exhibits a strong dependence on the separation distance $d$ between the 2DES and the gate. In the non-retarded limit, defined by $q_u \gg k_0$, and for a sufficiently small gate-to-2DES distance such that $q_u d \ll 1$, the reflection phase takes the value $\arg(r)/\pi \approx -\pi/4$ (see Fig. \ref{Fig-reflectance-ruu}(b)). This result contrasts with the prediction of the plane-wave matching method, which yields a reflection phase of $\pi$ under the same conditions. The nontrivial phase of plasmon reflection from the gate edge leads to an unexpected effect of changing the wave vector quantization rule for standing plasmon modes in the gap above a 2DES formed by two parallel metal gates, verified experimentally \cite{Accompanying}. The transmittance phase $arg(t_{ug})$ also depends on the degree of the 2DES screening, and can differ slightly from $\pi$ both upwards and downwards (Fig. \ref{Fig-transmittance-tug}(b)). Interestingly, for $arg(t_{ug})<\pi$ the transmission amplitude $|t_{ug}|>1$, while for $arg(t_{ug})>\pi$ the transmittance is less than unity ($|t_{ug}|<1$). The latter pattern holds true up to $q_ud \gg 1$ (see Fig. \ref{Fig-transmittance-tug}(a),(b))).

%коэффициент пропускания ведет себя также, как декремент моoности плазмона в структуре в виде волновода  с плотным диэлектриком в середине толщиной d.

\section{Results: incidence from gated section}

To solve the problem of scattering a plasmon propagating from the gate region toward the open region of the 2DES, we assume that the total electric field at an arbitrary point $z$ is determined by the gate current taking into account their screening by a 2DES. The gate current is decomposed into a superposition of an external current $J_{ext}$, associated with the plasmon in the gated 2DES, and a currents arising from the scattering of this plasmon at the gate edge $J_{\rm g, scat}$. Then, giving that the absence of the ungated plasmon propagating to gated region of 2DES (${E}_{\rm inc}=0$ in (\ref{Eq_total})), the equation (\ref{Eq_total}) can be rewritten as follows:

\begin{gather}
\label{Eq_total_by_gate_curr}
{{E}_{}}\left( q \right)={[J_{ext}+{J}_{\rm g, scat}}\left( q \right)]\frac{i}{M(q)}\frac{Z_0}{2 k_0}{{e}^{-\kappa(q)|z-d|}},
\end{gather}
where $J_{ext}=J_0/(q+q_g)$ is the Fourier transform of the gate current induced by gated plasmon incident from the right-half space. We write the equation (\ref{Eq_total_by_gate_curr}) in the gate plane ($z=d$), and using the Wiener-Hopf method obtain expressions for the scattered current  $J_{\rm g, scat}(q)$  in space $x>0$ and the electric field $E_L(q)$ at $x<0$ induced by currents in the gate.
\begin{gather}
\label{eq-scat-solution_gate_curr}
{{E}_{\rm L}}\left( q \right) = \frac{{{J}_0}}{q+{q_g}}\frac{Z_0}{2{{M}_{+}}\left( q \right){M}_{-}\left( -q_g \right)k_0}, \\ 
{{J}_{\rm g, scat}}\left( q \right) = -\frac{i{J_0}}{q+{{q}_g}}\frac{{{M}_{-}}\left(q\right)}{{M}_{-}\left( -q_g \right)}.
\end{gather}

Introducing the current $J_{g,scat}$ into equation (\ref{Eq_total_by_gate_curr}) allows us to obtain an expression for the total electric field at an arbitrary point $z$, caused by the currents in the gate:
\begin{multline}   
\label{eq-scat-solution_gate_curr}
{{E}_{\rm 2des}}\left( q \right) = \frac{{{J}_0}Z_0}{2k_0(q+{q_g})} \times
\\
\times \left[\frac{\kappa(-q_g)e^{-\kappa(-q_g)d}}{\varepsilon(-q_g)}-\frac{e^{-\kappa(q)d}}{{{M}_{+}}\left( q \right){M}_{-}( -q_g)\varepsilon_g(q)}\right].
\end{multline}

%The expression for electric field (\ref{Eq_total_by_gate_curr}) contains complete information about the field of the plasmons reflected under the gate and passed into the ungated region of the 2DES, as well as the radiation of bulk waves and evanescent fields occurring near the edge of the gate.

Equation (\ref{Eq_total_by_gate_curr}), considered at  $z=0$, allows us to obtain a simple relationship between the electric field in the 2DES $E_{0}$ and the source current $J_0$ in gate associated with the gated plasmon in 2DES:
\begin{equation}
\label{eq-E0-gated}
{E_{0}}=-\frac{J_0Z_0}{2\eta e^{-\kappa(q_g)}},
\end{equation}
than the amplitudes of reflected gated and transmitted ungated plasmon can be singled out from the total field (\ref{Eq_total_by_gate_curr}) by the residue of ${{E}_{\rm 2DES}}\left( q \right)$ at the poles $q=q_g$ and $q=-q_u$ timed by $i$, respectively:

\begin{gather}
r_{gg} E_0 =i\underset{q=q_g}{\text{Res}}E_{2des}{\rm}(q).\\
t_{gu} E_0 =i\underset{q=-q_u}{\text{Res}}E_{2des}{\rm}(q),
\end{gather} 

Taking into account (\ref{eq-E0-gated}), the reflection $r_{gg}$ and transmission $t_{gu}$ coefficients for a gated plasmon propagating  to the ungated 2DES have the form:

\begin{equation}
\label{eq-rgg}
{r_{gg}}=\frac{-i}{2q_g}\frac{\eta\kappa(q_g)}{k_0M_{+}(q_g)^2}\frac{e^{-2\kappa(q_g)d}}{{\left. \partial \varepsilon_g /\partial q \right|}_{q={q_g}}}.
\end{equation}
\begin{equation}
\label{eq-tgu}
{t_{gu}}=\frac{-i}{(q_g-q_u)}\frac{\eta M_{-}(-q_u)}{k_0M_{-}(-q_g)}\frac{\kappa(-q_u)e^{-[\kappa(q_g)+\kappa(q_u)]d}}{{\left. \partial \varepsilon_u /\partial q \right|}_{q=-{q_u}}}.
\end{equation}

\begin{figure}[ht!]
\center{\includegraphics[width=\linewidth]{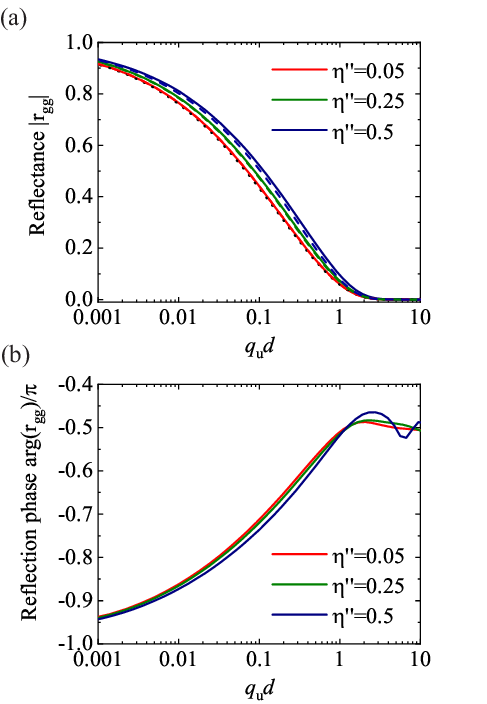}}
\caption{Reflection of the gated plasmon at a gate edge. Amplitude of the reflection coefficient $|r_{gg}|$ (solid curves) and $|r_{pw}|$ (dotted curve) (a) and its phase (b), both plotted as functions of the normalized gate-2DES separation $q_u d$. Different colors correspond to different values of normalized conductivity, $\eta ''$, which is assumed purely imaginary ($\eta'\ll \eta''$). Dashed curves at Fig. (a) correspond show the fully analytical approximation of the reflectance using absolute values of the factorized functions ~\eqref{eq_fact_analyt},~\eqref{eq_fact_analyt_g}.} 
\label{Fig-reflectance-rgg}
\end{figure}

The absolute reflectance $|r_{gg}|$ increases as the gate approaches 2DES and for $q_ud \ll 1$ strongly depends on the normalized conductivity 2DES $\eta''$ similarly to $|r_{uu}|$. The reflection calculated by the plane wave matching method $|r_{gg}^{pw}|$ (the black dotted curve at Fig.\ref{Fig-reflectance-rgg} (a)) is in good agreement with $|r_{gg}|$ only in the non-retarded limit, while the phase of reflection under the gate $arg(r_{gg})$ has a nontrivial dependence on the distance $d$, in contrast to $arg(r_{gg}^{pw})$=0 (Fig.\ref{Fig-reflectance-rgg}(b)). At close 2DES- gate separation arg($r_{gg}\approx-\pi$, while at
$q_ud\gg 1$ the gated plasmon reflection phase tends to $-\pi/2$.
%and $\pi$, which leads to the quantization rule $q_g=M\pi/L$, where $M=1,2,3...$ for the wavenumber of gated plasmons under a single gate of length $L$ (see the curve M=1 in Fig. 3 from the article PRL GaN???). 

\begin{figure}[ht!]
\center{\includegraphics[width=\linewidth]{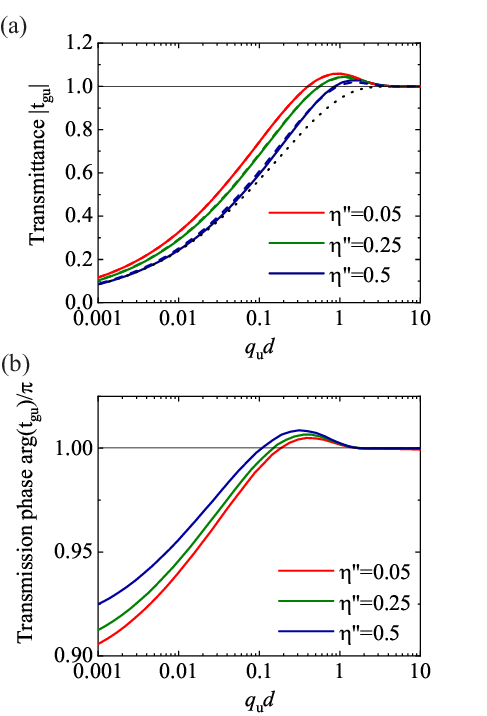}}

\caption{Transmission of the gated plasmon to ungated 2DES. Magnitude of the reflection coefficient $|t_{gu}|$ (solid curves) and $|t_{gu}^{pw}|$ (dotted curve) (a) and its phase (b), both plotted as functions of the normalized gate-2DES separation $q_u d$. Different colors correspond to different values of normalized conductivity $\eta$, which is assumed purely imaginary ($\eta' \ll \eta''$). Dashed curves at Fig. (a) correspond show the fully analytical approximation of the reflectance using absolute values of the factorized functions ~\eqref{eq_fact_analyt},~\eqref{eq_fact_analyt_g}. } 
\label{Fig-transmittance-tgu}
\end{figure}

In contrast to $|t_{ug}|$, the transmittance $|t_{gu}|$ decreases in limit $q_u d \to 0$; nevertheless, for certain values of $q_u d$ it may slightly exceed unity (Fig. \ref{Fig-transmittance-tgu}(a)). This dependence persists up to $q_u d \gg 0$ as long as the condition $\arg(t_{gu})/\pi > \pi$ remains satisfied (Fig. \ref{Fig-transmittance-tgu}(b)).

\section{Analysis of energy conservation and radiative decay}
\begin{figure}[ht]
\center{\includegraphics[width=\linewidth]{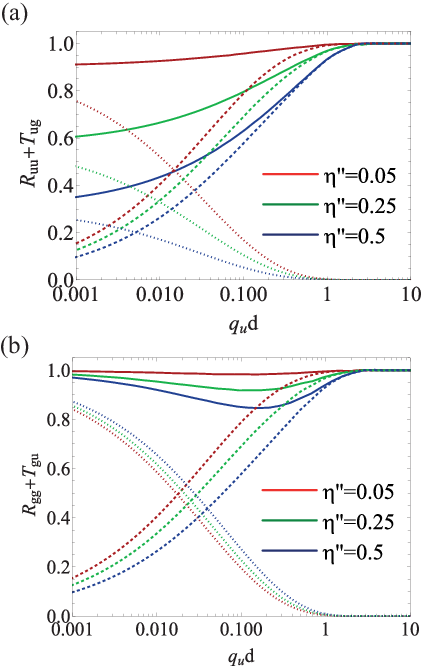}}

\caption{(a) The power reflectance $R_{uu}$ and transmittance $T_{ug}$ of ungated plasmon(a) and the power reflectance $R_{gg}$ and transmittance $T_{gu}$ of gated plasmon (b) both plotted as functions of the normalized gate-2DES separation $q_u d$. Sum of the power reflectance and transmittance shown by solid curves. Different colors correspond to different values of normalized conductivity $\eta$, which is assumed purely imaginary ($\eta' \ll \eta''$).} 
\label{Fig-R+T}
\end{figure}

Despite the fact that the transmittance amplitude of gated and ungated plasmons can exceed unity for some values of $q_u d$, the energy conservation law is satisfied for any $q_u d$, which corresponds to $R_{ii}+T_{ij}<1$ (Fig.~\ref{Fig-R+T}(a),(b)), where $R_{ii}=S_{ii}^{(r)}/S_{ii}^{(inc)}=|r_{ii}|^2$ is the reflection coefficient of ungated ($i=u$) and gated ($i=g$) plasmons, and $T_{ij}=S_{ij}^{(t)}/S_{i}^{(inc)}$ is the transmission coefficient of ungated ($i=u, j=g$) and gated ($i=g, j=u$) plasmons, where $S_{i}^{inc}$, $S_{ii}^{(r)}$ and $S_{ij}^{t}$ are the Poynting fluxes of the incident, reflected, and transmitted waves, respectively. Since we set the dissipative losses in 2DES to be extremely small ($\eta' \ll \eta''$), we can say that with increasing $\eta''$ the deviation of $R_{ii}+T_{ij}$ from unity is associated with the radiative losses caused by the emission of bulk waves as a result of plasmon scattering at the gate edge. The efficiency of plasmon radiation into bulk waves increases with increasing inductance of the 2DES, with the maximum radiation from scattering of an ungated plasmon occurring at a close gate-2DES distance (Fig.~\ref{Fig-R+T}(a)), whereas for gated plasmons there are an optimal values of $q_u d \approx \eta''$, when radiation losses are maximized (Fig.~\ref{Fig-R+T}(b)).

%Таким образом, показано, что в сильноэкранированных слабодиссипативных ДЭС полное поле  может быть описано одной плазмонной гармоникой, поскольку  

Although the expressions for the reflection \eqref{eq-ruu},\eqref{eq-rgg} and transmission \eqref{eq-tug}, \eqref{eq-tgu} are obtained analytically, the numerical implementation for factoring the dielectric functions \eqref{eq-factorisation} can present some difficulties. To determine the reflection and transmission amplitudes, we can write simplified expressions for the factorized dielectric functions, retaining only the terms that determine their absolute values. To achieve this, we expand the logarithm $\ln \varepsilon_{u,g}(q)$ in powers of small $\eta''$, and retain only the terms contributing to the absolute values of factorized functions. This results in
\textcolor{black}{
\begin{equation}
\label{eq_fact_analyt}
    |\varepsilon_{\pm}(q)|\approx \sqrt{|\varepsilon(q)|\left|\frac{q\pm q_{u}}{q \mp q_{u}}\right|} \exp\left(\pm\frac{\eta^{''}(q-|\kappa(q)|\text{sgn}(q)}{2\pi k_0}\right),
    \end{equation}
\begin{multline}
\label{eq_fact_analyt_g}
    |\varepsilon_{g\pm}(q)|\approx \sqrt{|\varepsilon_g(q)\left|\frac{q \pm q_{g}}{q \mp q_{g}}\right|}  \times \\    \exp\left(\pm\frac{z_0\eta^{''}}{\pi}\left[\frac{|\kappa(q)|}{k_0}\ln \left|\frac{\kappa_{-}(q)}{\kappa_{+}(q)}\right|-2q\right]\right).
   \end{multline}}
The approximate factorized 
 dielectric functions (\ref{eq_fact_analyt}),(\ref{eq_fact_analyt_g}) describe well the amplitudes of reflected and transmitted plasmons in a wide range of parameters. The values of $|r|$ and $|t|$, written using \eqref{eq_fact_analyt}, \eqref{eq_fact_analyt_g} are in good agreement with the full expressions for an arbitrary distance between the gate and 2DES, especially in the non-retarded limit (see dashed curves at Fig. \ref{Fig-reflectance-ruu}(a)-Fig. \ref{Fig-transmittance-tgu}(a)

\section{Discussion and conclusions}
In summary, we theoretically investigated the scattering of 2d- plasmons at the boundary between the gated and ungated regions of a 2DES. Using an exact solution of the scattering problem via the Wiener–Hopf method, we obtain analytical expressions for the complex reflection and transmission coefficients of plasmons incident on the boundary both from the ungated side and from under the gate. The derived expressions for the reflection and transmission coefficients account for the excitation of evanescent fields near the gate edge and the radiative losses due to emission of bulk electromagnetic waves in free space. In the nonretarded limit, the amplitude of the reflection coefficients $|r_{ii}|$, where $i=u,g$ tends to unity at small distances between the gate and the 2DES. This means that in the limit $d\to0$ the total electric field can be described by just a single reflected plasmon harmonic, for which the radiative losses is negligible, and the evanescent fields are very small compared to the plasmon field and do not contribute significantly to the total field. It is found that phases of the reflection and transmission coefficients exhibit nontrivial dependencies on the gate–2DES distance and on the 2DES conductivity, which fundamentally differ from the predictions of the plane-wave matching method that yields reflection and transmission phases equal to 0 or $\pi$. %The nontrivial phase shift can modify the quantization rule for the wave vectors of standing plasmon modes in slot cavities, as has been experimentally confirmed for ungated plasmon modes localized in the slot between two parallel metal gates [Murav].

The most efficient conversion of plasmons into bulk waves in free space upon scattering at the gate edge occurs at different structural parameters for gated and ungated plasmons: namely, at an ultra‑close gate ($q_ud\ll 1$) for scattering of an ungated plasmon, and at an optimal gate–2DES distance $d\approx \eta''/q_u$ for a gated plasmon.
Approximate expressions for the factorized dielectric functions \eqref{eq_fact_analyt}, \eqref{eq_fact_analyt_g} are proposed, which allow accurate calculation of the amplitudes of reflected and transmitted plasmons over a wide range of parameters, including the non‑retarded limit.
Thus, the developed theory provides a rigorous description of plasmon scattering at the gate–ungated interface and can serve as a foundation for modeling plasmonic crystals, including those for terahertz detection and amplification in structures based on two‑dimensional electron systems.

This work was supported by the Russian Science Foundation (Grant No. 24-79-00094). 
\bibliography{apssamp.bib}

\end{document}